\documentclass[12pt]{article}
\usepackage{amsmath,amssymb}
\usepackage{bm}
\usepackage{color}
\usepackage{graphicx}
\usepackage{cite}

\begin{document}

\title{
	\begin{flushright}
		\ \\*[-80pt]
		\begin{minipage}{0.2\linewidth}
			\normalsize
			%arXiv:YYMM.NNNN \\
			EPHOU-19-010\\
			HUPD1911 \\*[50pt]
		\end{minipage}
	\end{flushright}
	{\Large \bf
		New $A_4$ lepton flavor model from $S_4$ modular symmetry
		\\*[20pt]}}

\author{
	Tatsuo Kobayashi $^{1}$
	%\footnote{A's mail}
	,~Yusuke Shimizu $^{2}$
	%\footnote{B's mail}
	,~Kenta Takagi $^{2}$
	%\footnote{C's mail}
	,\\Morimitsu Tanimoto $^{3}$
	%\footnote{D's mail}
	,~Takuya H. Tatsuishi  $^{1}$
	%\footnote{D's mail}
	\\*[20pt]
	\centerline{
		\begin{minipage}{\linewidth}
			\begin{center}
				$^1${\it \normalsize
					Department of Physics, Hokkaido University, Sapporo 060-0810, Japan} \\*[5pt]
				$^2${\it \normalsize
					Graduate School of Science, Hiroshima University, Higashi-Hiroshima 739-8526}\\*[5pt]
				$^3${\it \normalsize
				Department of Physics, Niigata University, Niigata 950-2181}
			\end{center}
		\end{minipage}}
	\\*[50pt]}

\date{
	\centerline{\small \bf Abstract}
	\begin{minipage}{0.9\linewidth}
		\medskip
		\medskip
		\small
		We study a flavor model with $A_4$ symmetry which originates from $S_4$ modular group.
		In $S_4$ symmetry, $Z_2$ subgroup can be anomalous, and then $S_4$ can be violated to $A_4$.
		Starting with a $S_4$ symmetric Lagrangian at the tree level, the Lagrangian at the quantum level has only $A_4$ symmetry when $Z_2$ in $S_4$ is anomalous.
		We obtain modular forms of two singlets and a triplet representations of $A_4$ by decomposing $S_4$ modular forms into $A_4$ representations.
		We propose a new $A_4$ flavor model of leptons by using those $A_4$ modular forms.
		We succeed in constructing a viable neutrino mass matrix through the Weinberg operator for both normal hierarchy (NH) and inverted hierarchy (IH) of neutrino masses.
		Our predictions of the CP violating Dirac phase $\delta_{CP}$ and the mixing $\sin^2\theta_{23}$ depend on the sum of neutrino masses for NH.
	\end{minipage}
}

\begin{titlepage}
	\maketitle
	\thispagestyle{empty}
\end{titlepage}
\newpage
\section{Introduction}

The origin of the flavor structure is one of important issues in particle physics.
The recent development of the neutrino oscillation experiments provides us important clues to investigate the flavor physics.
Indeed, the neutrino oscillation experiments have presented two large flavor mixing angles, which is a contrast to the quark mixing angles.
In addition to the precise measurements of the flavor mixing angles of leptons,
the T2K and NO$\nu$A strongly indicate the CP violation in the neutrino oscillation \cite{Abe:2018wpn, NOvA:2018gge}.
We are in the era to develop the flavor theory of leptons with the observation of flavor mixing angles and CP violating phase.

It is interesting to impose non-Abelian discrete symmetries for flavors.
In the last twenty years, the studies of discrete symmetries for flavors have been developed through the precise observation of flavor mixing angles of leptons \cite{
	Altarelli:2010gt,Ishimori:2010au,Ishimori:2012zz,Hernandez:2012ra,
King:2013eh,King:2014nza,Tanimoto:2015nfa,King:2017guk,Petcov:2017ggy}.
Many models have been proposed by using the non-Abelian discrete groups $S_3$, $A_4$, $S_4$, $A_5$ and other groups with larger orders to explain the large neutrino mixing angles.
Among them, $A_4$ flavor symmetry is attractive because $A_4$ group is the minimal one including a triplet irreducible representation.
A triplet representation allows us to give a natural explanation of the existence of three families of leptons \cite{
Ma:2001dn,Babu:2002dz,Altarelli:2005yp,Altarelli:2005yx, Shimizu:2011xg,Petcov:2018snn,Kang:2018txu}.
However, a variety of models is so wide that it is difficult to obtain a clear evidence of the $A_4$ flavor symmetry.

Superstring theory is a promising candidate for the unified theory of all interactions including gravity and matter fields such as quarks and leptons as well as the Higgs field.
Superstring theory predicts six-dimensional compact space in addition to four-dimensional space-time.
Geometrical aspects, i.e. the size and shape of the compact space, are described by moduli parameters.
Gauge couplings and Yukawa couplings as well as higher order couplings in four-dimensional low-energy effective field theory depend on moduli parameters.
A geometrical symmetry of the six-dimensional compact space can be the origin of the flavor symmetry \footnote{
	It was shown that stringy selection rules in addition to geometrical symmetries lead to certain non-Abelian flavor symmetries~\cite{Kobayashi:2006wq,Kobayashi:2004ya,Ko:2007dz,Abe:2009vi}.}.

The torus compactification as well as the orbifold compactification has the modular symmetry $\Gamma$ \footnote{
	For example, zero-modes in the torus compactification with magnetic fluxes transform non-trivially under 
	the modular symmetry \cite{Kobayashi:2018rad}.}.
It is interesting that the modular symmetry includes $\Gamma_2 \simeq S_3$, $\Gamma_3 \simeq A_4$, $\Gamma_4 \simeq S_4$, $\Gamma_5 \simeq A_5$ as finite groups \cite{deAdelhartToorop:2011re}.
Inspired by these aspects, recently a new type of flavor models was proposed \cite{Feruglio:2017spp}.
In Ref.~\cite{Feruglio:2017spp}, the $A_4$ flavor symmetry is assumed as a finite group of the modular symmetry.
Three families of leptons are assigned to certain $A_4$ representations like conventional flavor models.
Furthermore, Yukawa couplings as well as Majorana masses are assumed to be modular forms which are functions of the modular parameter $\tau$ and they are non-trivial representations under $A_4$.
% The modular forms of the weight 2 are fundamental and their products provide modular forms of larger weights.
% For the $A_4$ modular symmetry,
We have a modular form of $A_4$ triplet with weight 2 \cite{Feruglio:2017spp}.
The flavor symmetry $A_4$ is broken when the value of the modular parameter $\tau$ is fixed.
It is noted that one can construct flavor models without flavon fields.

The modular forms of the weight 2 have been constructed for the $S_3$ doublet \cite{Kobayashi:2018vbk}, the $S_4$ triplet and doublet \cite{Penedo:2018nmg}, and the $A_5$ quintet and triplets \cite{Novichkov:2018nkm}, as well as the $\Delta(96)$ triplet and the $\Delta(384)$ triplet \cite{Kobayashi:2018bff}.
The modular forms of the weight 1 and higher weights are also given for $T'$ doublet \cite{Liu:2019khw}.
By use of these modular forms, new flavor models have been constructed \cite{
	Criado:2018thu,Kobayashi:2018scp,Novichkov:2018ovf,Ding:2019xna,deAnda:2018ecu,Novichkov:2018yse,Baur:2019kwi,deMedeirosVarzielas:2019cyj,Novichkov:2019sqv,Kobayashi:2018wkl,Okada:2018yrn,Nomura:2019jxj,Nomura:2019yft,Okada:2019xqk}.

Discrete symmetries can be anomalous \cite{Krauss:1988zc,Ibanez:1991hv,Banks:1991xj}.
Anomalies of non-Abelian symmetries were studied in \cite{Araki:2008ek}.
(See also \cite{Ishimori:2010au,Ishimori:2012zz}.)
The anomaly of the modular symmetry was also discussed \cite{Kariyazono:2019ehj}.
In the $S_4$ symmetry, the $Z_2$ subgroup can be anomalous and then $S_4$ can be violated to $A_4$.
The $A_5$ symmetry is always anomaly-free.
Both $S_3$ and $A_4$ can be anomalous, and then they can be violated to Abelian discrete symmetries.
Thus, the $S_4$ is unique among $S_3$, $A_4$, $S_4$, $A_5$ in the sense that it can be violated by anomalies to another non-Abelian symmetry, $A_4$.
Even starting with a $S_4$ symmetric Lagrangian at the tree level, the Lagrangian at the quantum level has only the $A_4$ symmetry when $Z_2$ subgroup of $S_4$ is anomalous.
Our purpose is to show such a possibility in a phenomenological viewpoint.
We decompose $S_4$ modular forms into $A_4$ representations.
Such modulus functions are different from the modular forms in $\Gamma_3$.
We propose a new $A_4$ flavor model with those $A_4$ modular forms,
which is much different from the typical modular $A_4$ models \cite{Feruglio:2017spp,Criado:2018thu,Kobayashi:2018scp}.

This paper is organized as follows.
In section 2, we give a brief review on the modular symmetry and the $S_4$ anomaly. 
In section 3, we present our model for lepton mass matrices.
In section 4, we show our numerical results for lepton mixing angles, the CP violating Dirac phase and neutrino masses.
Section 5 is devoted to a summary.
Relevant representations of $S_4$ and $A_4$ groups are presented in Appendix A.
We list the input data of neutrinos in Appendix B.

\section{Modular symmetry and $S_4$ anomaly}

\subsection{Modular forms}

We give a brief review on the modular symmetry and modular forms.
The torus compactification is the simplest compactification.
We consider a two-dimensional torus which can be constructed as a division of the two-dimensional real space $\mathbb{R}^2$ by a lattice $\Lambda$, i.e. $T^2 = \mathbb{R}^2/\Lambda$.
We use the complex coordinate on $\mathbb{R}^2$.
The lattice $\Lambda$ is spanned by two vectors, $\alpha_1=2\pi R$ and $\alpha_2 = 2 \pi R \tau$,
where $R$ is a real and $\tau$ is a complex modulus parameter.
The same lattice is spanned by the following lattice vectors,
\begin{equation}
	\left(
	\begin{array}{c}
		\alpha'_2 \\ \alpha'_1
	\end{array}\right)=
	\left(
	\begin{array}{cc}
		a & b \\
		c & d 
	\end{array}
	\right)
	\left(
	\begin{array}{c}
		\alpha_2 \\ \alpha_1
	\end{array}\right),
\end{equation}
where $a,b,c,d$ are integer with satisfying $ad-bc=1$.
That is, the $SL(2,\mathbb{Z})$ symmetry.
Under $SL(2,\mathbb{Z})$, the modulus parameter transforms 
\begin{equation}
	\tau \longrightarrow \tau' = \gamma \tau = \frac{a\tau + b}{c \tau + d}.
\end{equation}
This modular symmetry is generated by two elements, $S$ and $T$, 
which transform $\tau$ as
\begin{equation}
	S: \tau \longrightarrow -\frac{1}{\tau},\qquad 
	T: \tau \longrightarrow \tau +1.
	\label{symmetry}
\end{equation}
They satisfy the following algebraic relations,
\begin{equation}
	S^2=(ST)^3=\mathbb{I}.
\end{equation}

If we impose the algebraic relation $T^N =\mathbb{I}$, we obtain the finite groups $\Gamma_N$ for $N=2,3,4,5$, 
and these are isomorphic to $S_3,A_4,S_4,A_5$, respectively.
We define the congruence subgroups of level $N$ as 
\begin{equation}
	\Gamma(N) = \left\{ \left( 
	\begin{array}{cc}  a & b \\ c & d 
	\end{array}
	\right)  \in  SL(2,\mathbb{Z}),   \qquad  \left( 
	\begin{array}{cc}  a & b \\ c & d 
	\end{array}
	\right)   = \left( 
	\begin{array}{cc}  1 & 0 \\ 0 & 1 
	\end{array}
	\right)  \quad ({\rm mod}~N) \right\}.
\end{equation} 
%The quotient $\Gamma/\Gamma(N)$ is the above subgroup, $\Gamma_N = %\Gamma/\Gamma(N)$.
For $N=2$, we define $\bar\Gamma(2)\equiv \Gamma(2)/\{\mathbb{I},-\mathbb{I}\}$.
Since the element $-\mathbb{I}$ does not belong to $\Gamma(N)$
% while, since the element $-I$\UTF{0081} does not belong to $\Gamma(N)$,
for $N>2$, we have $\bar\Gamma(N)= \Gamma(N)$.
The quotient groups defined as
$\Gamma_N\equiv \bar \Gamma/\bar \Gamma(N)$
are finite modular groups.

Modular forms of weight $k$ are the holomorphic functions of $\tau$ and transform as
\begin{equation}
	f_i(\tau) \longrightarrow (c\tau +d)^k \rho(\gamma)_{ij}f_j(\gamma \tau),
\end{equation}
where $\rho(\gamma)_{ij}$ is a unitary matrix.
Also, matter fields $\phi^{(I)}$ with the modular weight $k_I$ transform
\begin{equation}
	(\phi^{(I)})_i(x) \longrightarrow (c\tau +d)^{k_I} \rho(\gamma)_{ij}(\phi^{(I)})_j(x),
\end{equation}
under the modular symmetry.

In Ref.~\cite{Penedo:2018nmg}, the modular form of the level $N=4$ for $\Gamma_4 \simeq S_4$ have been constructed with the Dedekind eta function, $\eta(\tau)$, 
\begin{equation}
	\eta(\tau) = q^{1/24} \prod_{n =1}^\infty (1-q^n)~,
\end{equation}
%%%%%%%%%%%%%%%%%%%%%%%%%%
where $q = e^{2 \pi i \tau}$.
The modular forms of the weight 2 are written by
\begin{align}
	\begin{aligned}
	Y_1(\tau) & = Y(1,1,\omega,\omega^2,\omega,\omega^2|\tau),    \\ 
	Y_2(\tau) & = Y(1,1,\omega^2,\omega,\omega^2,\omega|\tau),    \\
	Y_3(\tau) & = Y(1,-1,-1,-1,1,1|\tau),                         \\
	Y_4(\tau) & = Y(1,-1,-\omega^2,-\omega,\omega^2,\omega|\tau), \\
	Y_5(\tau) & = Y(1,-1,-\omega,-\omega^2,\omega,\omega^2|\tau), 
	\end{aligned}\label{eq:Y12345}
\end{align}
where $\omega = e^{2\pi i /3}$ and 
\begin{eqnarray}
	Y(a_1,a_2,a_3,a_4,a_5,a_6 | \tau) & =& 
	a_1 \frac{\eta'(\tau +1/2)}{\eta(\tau +1/2)} +4a_2 \frac{\eta'(4\tau )}{\eta(4\tau )} \nonumber \\
	& &+\frac14 \sum_{m=0}^3a_{m+3}  \frac{\eta'((\tau +m) /4)}{\eta((\tau +m)/4 )} .
\end{eqnarray}
These five modular forms are decomposed into the ${\bf 3}'$ and ${\bf 2}$ representations under $S_4$,
\begin{equation}
	Y_{S_4 {\bf 2}}(\tau) =\left(
	\begin{array}{c}
		Y_1(\tau) \\
		Y_2(\tau) 
	\end{array}
	\right), \qquad Y_{S_4 {\bf 3}'}(\tau) =\left(
	\begin{array}{c}
		Y_3(\tau) \\
		Y_4(\tau) \\
		Y_5(\tau) 
	\end{array}
	\right).
\end{equation}

The generators, $S$ and $T$, are represented on the above modular forms,
\begin{equation}
	\rho(S)=\left(
	\begin{array}{cc}
		0        & \omega \\
		\omega^2 & 0      
	\end{array}\right), \qquad
	\rho(T)=\left(
	\begin{array}{cc}
		0 & 1 \\
		1 & 0 
	\end{array}\right),
\end{equation}
for ${\bf 2}$, and 
\begin{equation}
	\rho(S)=-\frac13\left(
	\begin{array}{ccc}
		-1        & 2\omega^2 & 2 \omega  \\
		2\omega   & 2         & -\omega^2 \\
		2\omega^2 & -\omega   & 2         
	\end{array}\right), \qquad
	\rho(T)=
	-\frac13\left(
	\begin{array}{ccc}
		-1        & 2\omega   & 2 \omega^2 \\
		2\omega   & 2\omega^2 & -1         \\
		2\omega^2 & -1        & 2\omega    
	\end{array}\right),
\end{equation}
for ${\bf 3}'$.
The modular form of larger weights are obtained as products of $Y_{S_4{\bf 2} }(\tau)$ and $Y_{S_4{\bf 3}' }(\tau)$.
Other representations are shown in Appendix A.

\subsection{Anomaly}

A discrete symmetry can be anomalous.
Each element $g$ in a non-Abelian symmetry satisfies $g^N=1$, that is, the Abelian $Z_N$ symmetry. 
If all of such Abelian symmetries in a non-Abelian symmetry are anomaly-free, 
the whole non-Abelian symmetry is anomaly-free.
Otherwise, the non-Abelian symmetry is anomalous, and anomalous sub-group is violated.
Furthermore, each element $g$ is represented by a matrix $\rho(g)$.
If $\det \rho(g)=1$,  the corresponding $Z_N$ is always anomaly-free.
On the other hand, if $\det \rho(g) \neq 1$, the corresponding $Z_N$ symmetry can be anomalous.
See anomalies of non-Abelian symmetries \cite{Araki:2008ek,Ishimori:2010au,Ishimori:2012zz}.

In particular, in Refs.~ \cite{Ishimori:2010au,Ishimori:2012zz}, 
it shows which sub-groups can be anomalous in non-Abelian discrete symmetries.
The $S_4$ group is isomorphic to $(Z_2 \times Z_2) \rtimes S_3$, and then 
the $Z_2$ symmetry of $S_3$ can be anomalous in $S_4$.
In general, the ${\bf 2}$ and ${\bf 3}$ representations as well as ${\bf 1}'$  have $\det \rho(g) =-1$ 
while the ${\bf 1}$ and ${\bf 3}'$ representations have $\det \rho(g)=1$.
Indeed $\rho(S)$ and $\rho(T)$ for ${\bf 2}$ as well as ${\bf 3}$ and ${\bf 1}'$ have $\det (\rho(S))=\det(\rho(T))=-1$.
Thus, the odd number of ${\bf 2}$'s as well as  ${\bf 3}$ and ${\bf 1}'$ can lead to anomalies. 

If the above $Z_2$ symmetry in $S_4$ is anomalous, $S_4$ is violated to $A_4$.
In this case, $S$ and $T$ themselves are anomalous, but $\tilde S = T^2$ and $\tilde T = ST$ are anomaly-free.
These anomaly-free elements satisfy 
\begin{equation}
	(\tilde S)^2 = (\tilde S \tilde T)^3 = (\tilde T)^3 = \mathbb{I},
\end{equation}
if we impose $T^4 = \mathbb{I}$.
That is, the $A_4$ algebra is realized.
The explicit representations of generators $\tilde S$ and $\tilde T$ for the  $A_4$ triplet and singlets are presented in Appendix A.
The modular forms for $S_4$ act under the $A_4$ symmetry as follows:
\begin{equation}
	Y_{S_4{\bf 2}}(\tau)\rightarrow
	(\ Y_{A_4{\bf 1''} }(\tau), \ Y_{A_4{\bf 1'}}(\tau)\ ) \ , \qquad 
	Y_{S_4{\bf 3}' }(\tau) \rightarrow Y_{A_4{\bf 3} }(\tau) \ .
\end{equation}
That is, we have 
\begin{equation}
	Y_{A_4{\bf 1}'}(\tau)=Y_{2}(\tau) ,
	\qquad Y_{A_4{\bf 1}''}(\tau)=Y_{1}(\tau), 
	\qquad
	Y_{A_4{\bf 3} }(\tau)= 
	\begin{pmatrix}
		Y_3 (\tau) \\
		Y_4 (\tau) \\
		Y_5(\tau)  
	\end{pmatrix}.
	\label{modularA4}
\end{equation}

%\begin{equation}
%	Y_{S_4{\bf 2}}(\tau)\rightarrow
%	(\ Y_{A_4{\bf 1'} }(\tau), \ Y_{A_4{\bf 1''}}(\tau)\ ) \ , \qquad 
%	Y_{S_4{\bf 3}' }(\tau) \rightarrow Y_{A_4{\bf 3} }(\tau) \ .
%\end{equation}
%That is, we have 
%\begin{equation}
%	Y_{A_4{\bf 1}'}(\tau)=Y_1(\tau) ,
%	 \qquad Y_{A_4{\bf 1}''}(\tau)=Y_{2}(\tau), 
%	\qquad
%	Y_{A_4{\bf 3} }(\tau)= 
%	\begin{pmatrix}
%		Y_3 (\tau) \\
%		Y_4 (\tau) \\
%		Y_5(\tau) 
%	\end{pmatrix}.
%\label{modularA4}
%\end{equation}

Note that these are different from modular forms of the level $N=3$ for $\Gamma_3$ 
because they do not transform as $A_4$ multiplets under $S$ and $T$.

%Note that these are not modular forms of the level $N=3$ for $\Gamma_3$  
%because $\tilde S =T^2$ and $\tilde T= ST$ do not generate $SL(2,\mathbb{Z})$ 
%for the $SL(2,\mathbb{Z})$ generators, $S$ and $T$ without imposing %$T^N=\mathbb{I}$.

Anomalies of the $S_4$ symmetry, in particular its $Z_2$ sub-symmetry,
depend on models, that is, the numbers of ${\bf 2}$, ${\bf 3}$ 
and ${\bf 1}'$.
If the $S_4$ symmetry is anomaly-free and exact, 
the model building follows the study in Ref.~\cite{Penedo:2018nmg} and its extension.
If the $S_4$ is anomalous and violated to $A_4$, that leads to a new type of model building.
In the next section, we study such a new possibility for lepton mass matrices.

\section{$A_4$ lepton model from  $S_4$ modular symmetry}

We present a viable $A_4$ model of leptons originated from the subgroup of $S_4$ group.
The charge assignment of the fields and modular forms is summarized in Table \ref{tb:charge}.
We assign the modular weight $-1$ to the left- and right-handed leptons.
If the $S_4$ is exact, $\mu^c_{\mathbf{1''}} $ and $\tau^c_{\mathbf{1'}}$ are 
combined to  the $S_4$ doublet.
The odd number of doublets can lead to anomalies.

%%%%%%%%%%%%%%%%%%%%%%%%%%%%%%%%%%%%%%%%%%
%The charge assignment of the fields and modular forms is summarized in Table \ref{tb:charge}.
%The parameters $\alpha, \beta, \gamma$ are determined by the observed %charged lepton masses and  the value of  $\tau$.
% --- Table ---%
\begin{table}[h]
	\centering
	\begin{tabular}{|c||c|c|c|c|c|c|} \hline
		        & $L_{\mathbf{3}}$ & $e^c_{\mathbf{1}},\mu^c_{\mathbf{1''}},\tau^c_{\mathbf{1'}}$ & $H_{u,d}$ & $Y_{A_4\mathbf{3}}$ & $Y_{A_4\mathbf{1'}}$ & $Y_{A_4\mathbf{1''}}$ \\ \hline \hline
		\rule[14pt]{0pt}{0pt}
		$SU(2)$ & $2$              & $1$                                                          & $2$       & $1$                 & $1$                  & $1$                   \\
		$A_4$   & $3$              & $1\, ,1''\, ,1'$                                             & $1$       & $3$                 & $1'$                 & $1''$                 \\
		$-k_I$  & $-1$             & $-1$                                                         & $0$       & $k = 2$             & $k = 2$              & $k = 2$               \\ \hline
	\end{tabular}
	\caption{
	The charge assignment of $SU(2)$, $A_4$, and the modular weight ($-k_I$ for fields and $k$ for coupling $Y$).}
	\label{tb:charge}
\end{table}

The modular forms of weight 2 that transform non-trivially under the $A_4$ symmetry
are given in $S_4$ modular group as discussed in section 2.
The $A_4$ triplet $Y_{A_4\bf 3}$ and non-trivial $A_4$ singlets $Y_{A_4\bf 1'}$, $Y_{A_4\bf 1''}$
are constructed by five modular forms in Eq.~\eqref{eq:Y12345},
which is a difference from the $\Gamma_3\simeq A_4$ modular symmetry with three modular forms.

Suppose neutrinos to be Majorana particles.
The superpotential of the neutrino mass term is given by the Weinberg operator:
\begin{align}
	\label{eq:wnu}                                                                                          
	w_\nu = \frac{1}{\Lambda}\bigg[ Y_{A_4\mathbf{3}} + a Y_{A_4\mathbf{1''}} + b Y_{A_4\mathbf{1'}} \bigg] 
	L_{\mathbf{3}} L_{\mathbf{3}} H_u H_u,                                                                  
\end{align}
where $L_{\bf 3}$ denote the $A_4$ triplet of the left-handed lepton doublet, $(L_e,L_\mu,L_\tau)^T$, and $H_u$ stands for the Higgs doublet which couples to the neutrino sector.
Parameters $a$ and $b$ are complex constants in general.
If the $S_4$ symmetry is exact, $Y_{A_4\mathbf{1'}} $ and $Y_{A_4\mathbf{1''}}$ are combined to 
the $S_4$ doublet  $Y_{S_4{\bf 2}}$.
That is, the second and third terms are originated 
from $aY_{S_4{\bf 2}}LLH_uH_u$, where $L$ is taken to  be
$\mathbf{3}'$ of $S_4$,
and we have $a=b$.
Breaking of $S_4$ to $A_4$ leads to the above terms with $a \neq b$.
One naively expects to be $a\sim b$, although their difference depends on breaking effects.
At any rate, we treat them as independent parameters from the phenomenological viewpoint.
We also discuss the situation with $a \sim b$.
%since $S_4$ is broken due to quantum effects.

The superpotential of the mass term of charged leptons is described as
\begin{align}
	\label{eq:we}                                                                                          
	w_e = 	\bigg[ \alpha e^c_{\mathbf{1}} + \beta \mu^c_{\mathbf{1''}} + \gamma \tau^c_{\mathbf{1'}}\bigg] 
	Y_{A_4\mathbf{3}} L_{\mathbf{3}}  H_d,                                                                 
\end{align}
where charged leptons $e^c_{\mathbf{1}}, \mu^c_{\mathbf{1''}}, \tau^c_{\mathbf{1'}}$ are assigned to the $A_4$ singlets of
${\bf 1, 1'', 1'}$ respectively.
The $H_d$ is a Higgs doublet which couples to the charged lepton sector.
Coefficients $\alpha$,   $\beta$ and  $\gamma$ 
can be taken  to be real.
Then, charged lepton masses are given in terms of $\tau$,
$\langle H_d \rangle$, $\alpha$,   $\beta$ and  $\gamma$. 
Similar to Eq.~(\ref{eq:wnu}), if the $S_4$ is exact, $\mu^c_{\mathbf{1''}} $ and $\tau^c_{\mathbf{1'}}$ are 
combined to  the $S_4$ doublet.
That is, we have to require $\beta = \gamma$.
Here, we also treat these parameters  as independent parameters from the phenomenological viewpoint.
%We also expect $\beta\sim \gamma$ 
%since $S_4$ is broken due to quantum effects.

% -------------%
%
%
%
The relevant mass matrices are given by using  
the multiplication rules based on  $\tilde S$ and $\tilde T$
in Appendix A. 
The Majorana neutrino mass matrix is:
\begin{align}
	M_\nu = \frac{\langle H_u \rangle ^2}{\Lambda}\left[
	\begin{pmatrix}
	2Y_3 & -Y_5 & -Y_4 \\
	-Y_5 & 2Y_4 & -Y_3 \\
	-Y_4 & -Y_3 & 2Y_5 
	\end{pmatrix}
	+ a Y_{1}
	\begin{pmatrix}
	0    & 1    & 0    \\
	1    & 0    & 0    \\
	0    & 0    & 1    
	\end{pmatrix}
	+ bY_{2}
	\begin{pmatrix}
	0    & 0    & 1    \\
	0    & 1    & 0    \\
	1    & 0    & 0    
	\end{pmatrix}  \right],
	\label{neutrino}
\end{align}
while the charged lepton matrix is given as:
\begin{align}
	M_e = \langle H_d \rangle \begin{pmatrix}
	\alpha & 0     & 0      \\
	0      & \beta & 0      \\
	0      & 0     & \gamma 
	\end{pmatrix}
	\begin{pmatrix}
	Y_3    & Y_5   & Y_4    \\
	Y_4    & Y_3   & Y_5    \\
	Y_5    & Y_4   & Y_3    
	\end{pmatrix}_{RL},
	\label{lepton}
\end{align}
where $\alpha$, $\beta$ and $\gamma$ are taken to be  real positive without loss of generality.

\section{Numerical result}

\begin{figure}[b!]
	\begin{tabular}{ccc}
		\begin{minipage}{0.5\hsize}                                                                                             
		\includegraphics[bb=0 0 400 280,width=\linewidth]{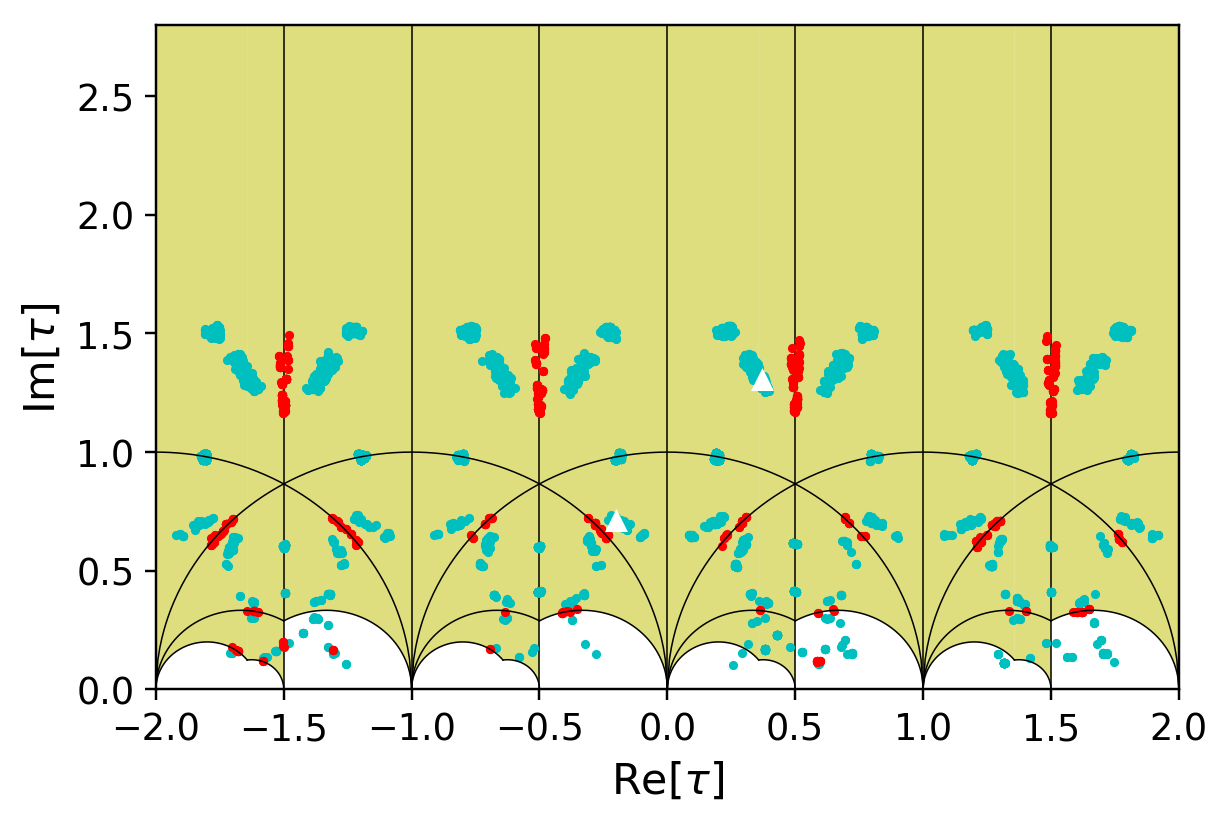}                                                      
		\caption{ Allowed  region on the	${\rm Re}[\tau]$--$ {\rm Im}[\tau]$ plane.                                             
		The fundamental domain of $\Gamma(4)$ are shown by olive-green.                                                         
		Cyan-points and red-points denote cases of NH and IH, respectively. Two small white triangles  denote a pair connecting 
		by the $S$ symmetry.}                                                                                                   
		\label{fig:tau}                                                                                                         
		\end{minipage}                                                                                                          
		\phantom{=}                                                                                                             
		\begin{minipage}{0.5\hsize}                                                                                             
		\includegraphics[bb=0 0 400 280,width=\linewidth]{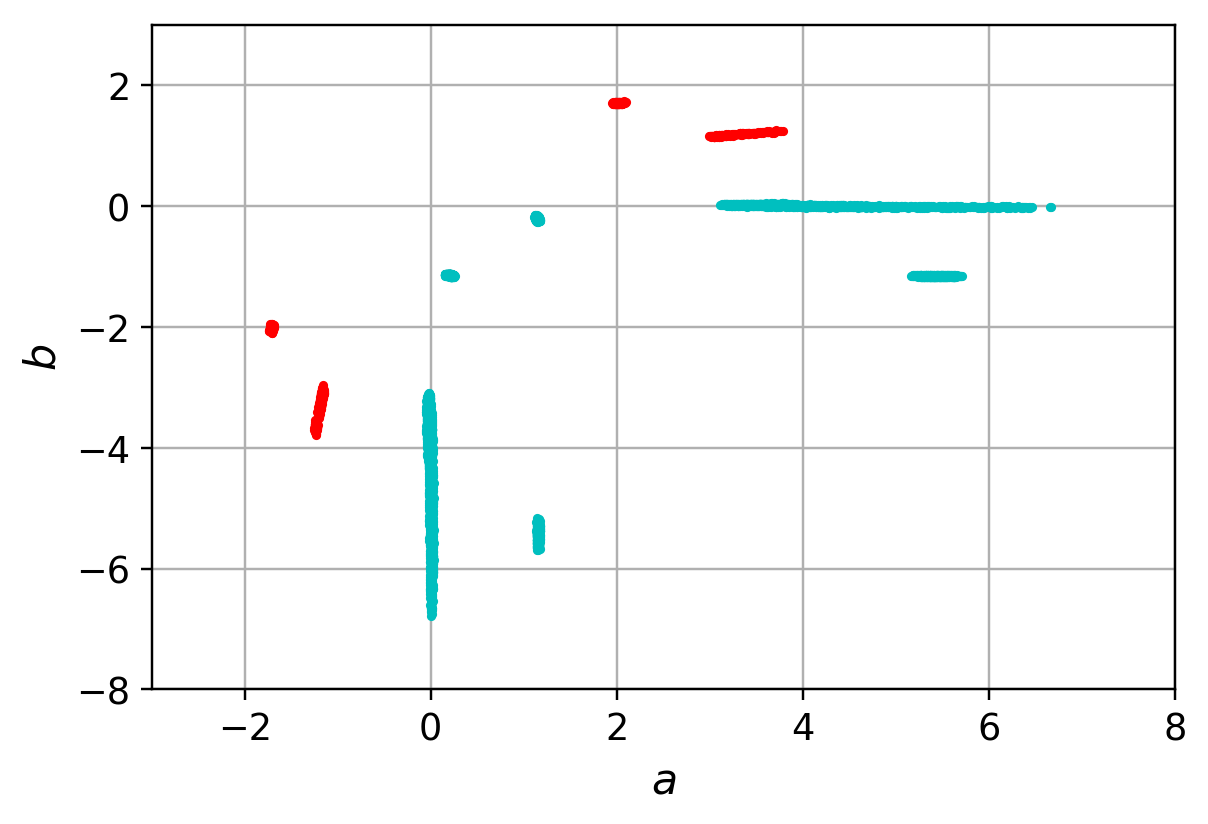}                                                              
		\caption{Allowed region on the  $a$--$b$ plane, where $a$ and $b$ are taken to be real.                                 
		Cyan-points and red-points denote cases of NH and IH, respectively.}                                                    
		\label{fig:ab}                                                                                                          
		\end{minipage}                                                                                                          
	\end{tabular}
\end{figure}

We discuss numerical results for the lepton flavor mixing by using Eqs.\,(\ref{neutrino}) and (\ref{lepton}).
Parameters of the model are
$\alpha$, $\beta$, and  $\gamma$ of the charge lepton mass matrix;
and $a$ and $b$ of the neutrino mass matrix in addition to modulus $\tau$.
Parameters  $\alpha$, $\beta$, and  $\gamma$ are real while
$a$ and $b$ are complex in general.
However, we take $a$ and $b$ to be real in order to present
a simple viable model, that is to say, the CP violation comes from modular forms in section 2.
Parameters $\alpha$, $\beta$, and  $\gamma$ are given in terms of $\tau$ after inputting three charged lepton masses.
Therefore, we scan the parameters in the following ranges as:
\begin{align}
	\tau = [-2.0, 2.0] + i [0.1, 2.8], \quad 
	a = [-15, 15], \quad                     
	b = [-15, 15],                           
\end{align}
where the fundamental domain of $\Gamma(4)$ is taken into account.
The fundamental region is shown in Fig.~\ref{fig:tau}.
The lower-cut $0.1$ of ${\rm Im[\tau]}$ is artificial to keep the accurate numerical calculation.
The upper-cut $2.8$ is enough large to estimate the modular forms.

We input the experimental data within $3\,\sigma$ C.L.\cite{Esteban:2018azc} of three mixing angles in the lepton mixing matrix \cite{Tanabashi:2018oca} in order to constrain magnitudes of parameters.
We also put the two observed  neutrino mass square differences
($\Delta m_{\rm sol}^2$, $\Delta m_{\rm atm}^2$)
and the cosmological bound for the neutrino masses $\sum m_i < 0.12$ [eV]~\cite{Vagnozzi:2017ovm,Aghanim:2018eyx}.
Since parameters are severely restricted due to experimental data, the Dirac phase $\delta_{CP}$ is predicted.
Furthermore, we also discuss the effective mass of the $0\nu\beta\beta$ decay $\langle m_{ee}\rangle$:
\begin{align}
	\langle m_{ee}	\rangle=\left| m_1 c_{12}^2 c_{13}^2+ m_2s_{12}^2 c_{13}^2 e^{i\alpha_{21}}+ 
	m_3 s_{13}^2 e^{i(\alpha_{31}-2\delta_{CP})}\right|  \ ,                                    
\end{align}
where $\alpha_{21}$ and $\alpha_{31}$ are Majorana phases defined in Ref. \cite{Tanabashi:2018oca}.

%%%%%%%%%%%%%%%%%%%%%%%%%%%%%%%%%%%%%%%%%%%%%%%%%%

\begin{figure}[b!]
	\begin{tabular}{ccc}
		\begin{minipage}{0.49\hsize}                                                         
		\includegraphics[bb=0 0 400 280,width=\linewidth]{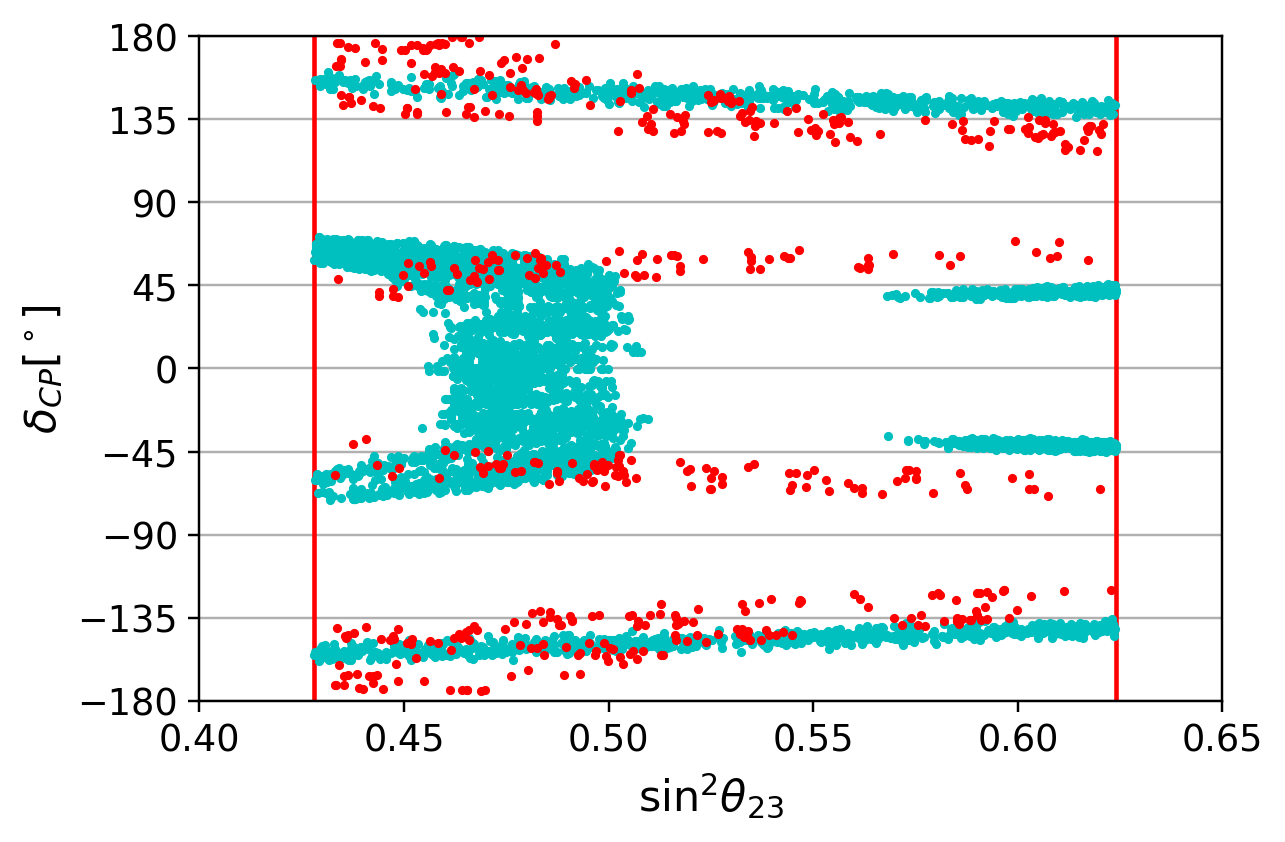}                      
		\caption{Predicted $\delta_{CP}$ versus $\sin^2\theta_{23}$, where                   
		cyan-points and red-points denote cases of NH and IH, respectively.                  
		The vertical red lines denote  $3\sigma$ interval of data.	}                         
		\label{fig:delta-23}                                                                 
		\end{minipage}                                                                       
		\phantom{=}                                                                          
		\begin{minipage}{0.49\hsize}                                                         
		\includegraphics[bb=0 0 400 280,width=\linewidth]{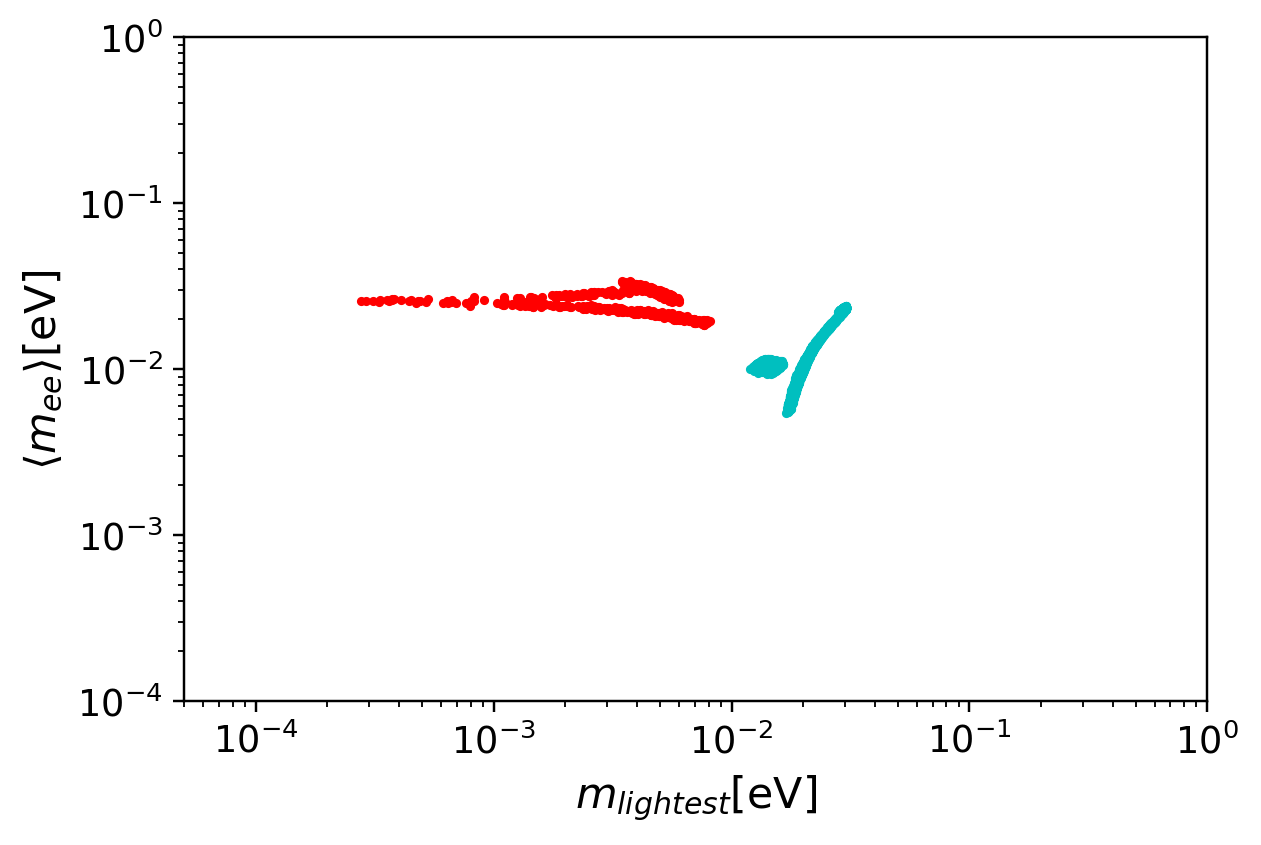}                        
		\caption{Predicted $\langle m_{ee} \rangle$ versus the lightest neutrino mass, where 
		cyan-points and red-points denote cases of NH and IH, respectively.                  
		The cosmological bound of $\sum m_i$ is imposed.}                                    
		\label{fig:mee1}                                                                     
		\end{minipage}                                                                       
	\end{tabular}
\end{figure}

There are two possible spectra of neutrinos masses $m_i$,
which are the normal hierarchy (NH), $m_3>m_2>m_1$, and the inverted hierarchy (IH), $m_2>m_1>m_3$.
At first, we show the predicted region of $\sin^2\theta_{23}$--$\delta_{CP}$ in Fig.\,\ref{fig:delta-23}, 
where cyan-points and red-points denote cases of NH and IH, respectively.
%The correlation between $\sin^2\theta_{23}$ and $\delta_{CP}$ is found %distinctly.
For NH of neutrino masses, the predicted $\delta_{CP}$ is
$|\delta_{CP}|<70^\circ$ and $|\delta_{CP}|=135^\circ$--$160^\circ$.
It is noticed that $|\delta_{CP}|\simeq 90^\circ$ is excluded.
The prediction of $\delta_{CP}$ becomes clear if $\sin^2\theta_{23}$ is precisely measured.
Indeed, $\delta_{CP}$ is predicted around
$\pm 40^\circ$ and $\pm 140^\circ$ at the observed best fit point of $\sin^2\theta_{23}=0.582$ \cite{Esteban:2018azc}.

For IH of neutrino masses, the predicted $\delta_{CP}$ is
$|\delta_{CP}|=40^\circ$--$70^\circ$ and $|\delta_{CP}|=110^\circ$--$180^\circ$.
It is found that $|\delta_{CP}|\simeq 90^\circ$ is also excluded for IH. 

%%%%%%%%%%%%%%%%%%%%%%%%%%%%%%%%%%%%%%%%%%%%%%%%%%%%

%%%%%%%%%%%%%%%%%%%%%%%%%%%%%%%%%%%%%%%%%%%%
%%%%%%%%%%%%%%%%%%%%%%%%%%%%%%%%%%%%%%%%%%%%%%%%%%%%%%
We present the prediction of the effective mass of the $0\nu\beta\beta$ decay, $\langle m_{ee}\rangle$
versus the lightest neutrino mass for both NH and IH of neutrino masses in Fig.\,\ref{fig:mee1}.
The upper-bound of the lightest neutrino mass is given by the cosmological upper-bound of the sum of neutrino masses.
For NH, the lower-bound of the lightest neutrino mass is $12$ [meV].
The predicted range of  $\langle m_{ee} \rangle$ is  $5$--$22$ [meV] depending on the lightest neutrino mass.
For IH, $\langle m_{ee} \rangle$ is predicted in $15$--$30$ [meV].
Hence, the $0\nu\beta\beta$ decay will be possibly observed in the future \cite{Gando:2019mxj}.

%%%%%%%%%%%%%%%%%%%%%%%%%%%%%%%%%%%%%%%%%%%%
\begin{figure}[h]
	\begin{tabular}{ccc}
		\begin{minipage}{0.485\hsize}                                             
		\includegraphics[bb=0 0 400 280,width=\linewidth]{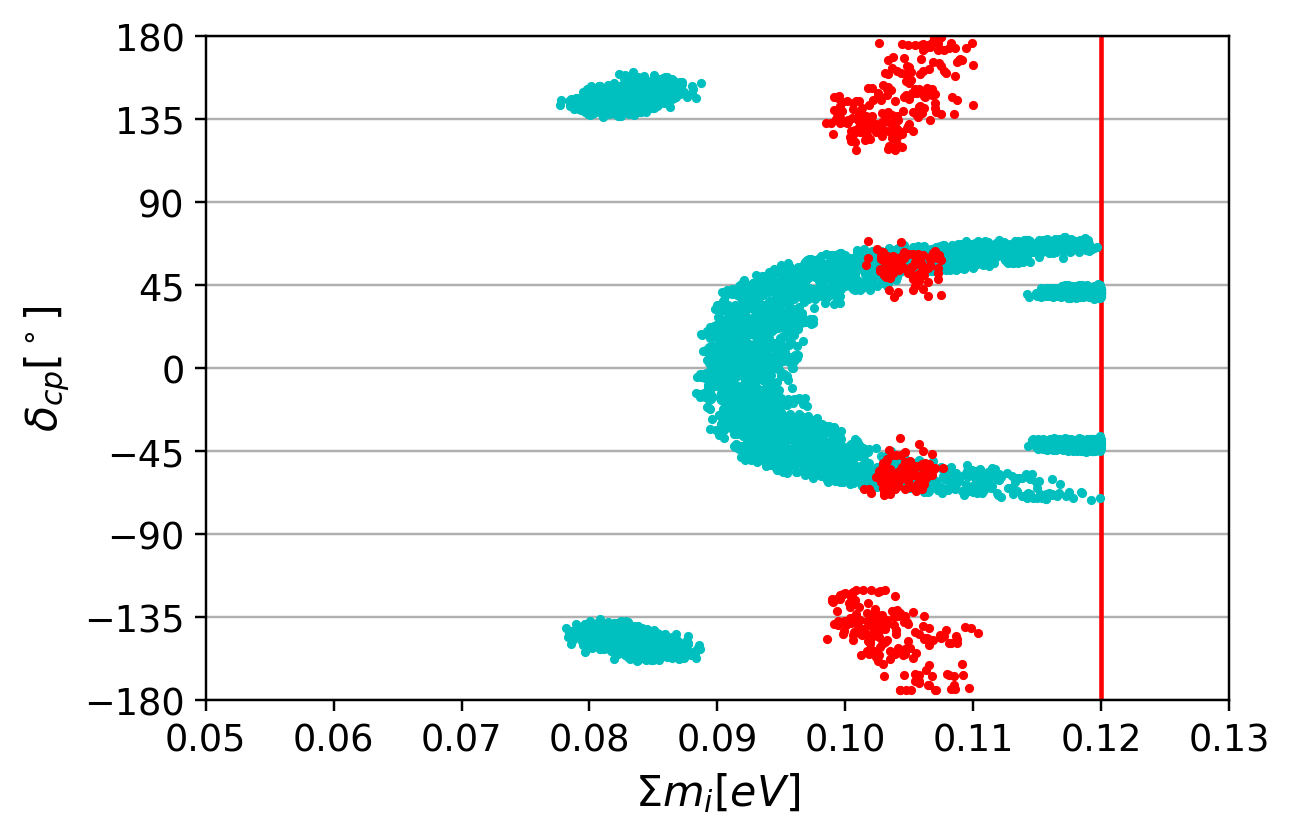}         
		\caption{Predicted $\delta_{CP}$ versus $\sum m_i$,                       
		where cyan-points and red-points denote cases of NH and IH, respectively. 
		The vertical red line denotes the cosmological upper-bound.}              
		\label{fig:delta-msum}                                                    
		\end{minipage}                                                            
		\phantom{=}                                                               
		\begin{minipage}{0.485\hsize}                                             
		\includegraphics[bb=0 0 400 280,width=\linewidth]{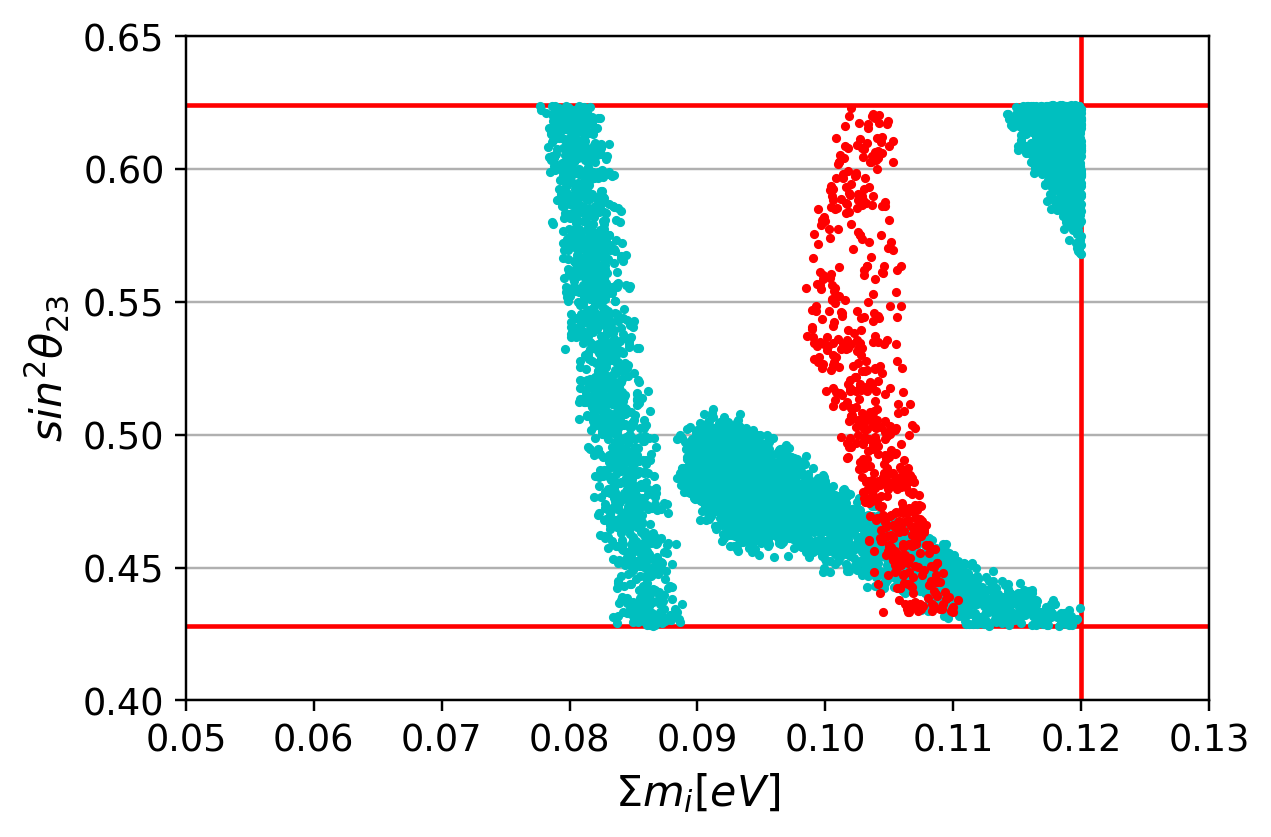}            
		\caption{Predicted $\sin^2\theta_{23}$ versus $\sum m_i$.                 
		The notation is the same as in Fig.\ref{fig:delta-msum}.                  
		Horizontal  red lines denote $3\sigma$ interval of                        
		the experimental  data. }                                                 
		\label{fig:23-msum}                                                       
		\end{minipage}                                                            
	\end{tabular}
\end{figure}
%%%%%%%%%%%%%%%%%%%%%%%%%%%%%%%%%%%%%%%%%%%%

%%%%%%%%%%%%%%%%%%%%%%%%%%%%%%%%%%%%%%%%%%%%
Let us discuss the neutrino mass dependence of $\delta_{CP}$ and $\sin^2\theta_{23}$.
We present the predicted $\delta_{CP}$ versus the sum of neutrino masses $\sum m_i$ in Fig.\,\ref{fig:delta-msum},
where the cosmological bound $\sum m_i < 120$ [meV] is imposed.
The predicted $\delta_{CP}$ depends on the sum of neutrino masses, where $\sum m_i > 78$ [meV] for NH of neutrino masses.
In the range of $78<\sum m_i < 88$ [meV], $\delta_{CP}\simeq \pm (
135^\circ$--$160^\circ)$ is predicted.
In the range of $\sum m_i>88$ [meV],
we obtain  $|\delta_{CP}|<70^\circ)$.
For IH, the sum of neutrino mass is predicted for $ 98 \text{ [meV]}< \sum m_i < 110 \text{ [meV]}$ with $|\delta_{CP}|>110^\circ$ or $40^\circ < |\delta_{CP}| < 70^\circ$.

The predicted $\sin^2\theta_{23}$ is also presented versus $\sum m_i$ in Fig.\,\ref{fig:23-msum}.
In the case of NH, 
the observed  best fit point of $\sin^2\theta_{23}=0.582$ \cite{Esteban:2018azc} is realized 
at $\sum m_i =80$--$85$ [meV]. 
%the predicted $\sin^2\theta_{23}$ is larger than $0.45$
%for  $\sum m_i\geq 98$ [meV].
For IH, we get $\sum m_i =100$--$ 105$ [meV] for the best fit point of $\sin^2\theta_{23}=0.582$.
Hence, the observation of the sum of neutrino masses in the cosmology will provide a severe constraint to the flavor model.
 
%%%%%%%%%%%%%%%%%%%%%%%%%%%%%%%%%%%%%%%%%%
We present the set of best-fit parameters and  observables.
For NH, we obtain: 
\begin{eqnarray}
	&&\tau=-1.717 + 0.5852  i, \quad  a= 0.2178, \quad  b= - 1.141, 
	\nonumber\\
	&&\alpha v_d = 1.73 \times 10^5{\rm eV}, \quad 
	\beta v_d =  4.64 \times  10^8{\rm eV}, \quad
	\gamma v_d = 3.34 \times 10^7{\rm eV},\label{sample-NH}\\
	&&\sin^2 \theta_{12} = 0.299, \quad
	\sin^2\theta_{23} = 0.587,\quad
	\sin^2 \theta_{13} = 0.0228, \quad \delta_{CP} = -142.9^\circ ,
	\nonumber\\
	&& \Delta m^2_{21} = 7.38 \times 10^{-5}{\rm eV^2}, \nonumber \
	\Delta m^2_{31} = 2.54 \times 10^{-3}{\rm eV^2}\nonumber, \
	\langle m_{ee}\rangle  = 13.4 {\rm meV}, \ \sum m_i= 81.5{\rm meV},
\end{eqnarray}
where  $\chi^2=0.31$.
For IH, we have:
\begin{eqnarray}
	&&\tau= -1.508 + 1.288 i, \quad  a = -1.230, \quad  b= -3.616,
	\nonumber\\
	&&\alpha v_d = 7.01 \times 10^7{\rm eV}, \quad 
	\beta v_d = 1.04 \times  10^9{\rm eV}, \quad
	\gamma v_d = 3.57 \times 10^5{\rm eV},\label{sample-IH}\\
	&&\sin^2 \theta_{12} =  0.292, \quad
	\sin^2\theta_{23} =  0.586,\quad
	\sin^2 \theta_{13} = 0.0227, \quad \delta_{CP} =  -133.4^\circ ,
	\nonumber\\
	&& \Delta m^2_{21} = 7.16\times 10^{-5}, \nonumber \quad
	\Delta m^2_{31} =-2.51 \times 10^{-3}\nonumber, \quad 
	\langle m_{ee}\rangle  = 27.7 {\rm meV},\quad \sum m_i= 101.5{\rm meV},
\end{eqnarray}
where  $\chi^2=0.70$.

%%%%%%%%%%%%%%%%%%%%%%%%%%%%%%%%%%%%%%%%%%%%
%%%%%%%%%%%%%%%%%%%%%%%%%%%%%%%%%%%%%%%%%%%%
%%%%%%%%%%%%%%%%%%%%%%%%%%%%%%%%%%%%%%%%%%
We show the allowed region of ${\rm Re}[\tau]$--$ {\rm Im}[\tau]$ in Fig.\,\ref{fig:tau}, 
where cyan-points and red-points denote the NH and IH cases, respectively.
The fundamental domain of $\Gamma(4)$ is also presented by olive-green in this figure,
where the real part of $\tau$ is $[-2,2]$ and the imaginary part of $\tau$ is expanded downward.
Some points are outside of the fundamental domain of $\Gamma(4)$.
Those points are transformed into the inside of the fundamental domain by the $S_4$ transformations.
%%%%%%%%%%%%%%%%%%%%%%%%%%%%%%%%%%%%%
%Many allowed points lie in the range ${\rm Im}[\tau] =1.2\text{--}1.5$.
%However, there are also allowed points of ${\rm Im}[\tau]$ considerably
% smaller than $1$, which contrasts with other models
% with the modular symmetry.
%%%%%%%%%%%%%%%%%%%%%%%%%%%%%%%%%%%%%
In this figure, the $\tau\to \tau+1$ shift symmetry of Eq.(\ref{symmetry}) is clearly seen.
In order to show $\tau\to -1/\tau$ symmetry of Eq.(\ref{symmetry}),
we plot one pair by small white triangles.
It is seen that the plotted points (red) on ${\rm Re}[\tau] =0.5$ inside the fundamental region of
$SL(2,\mathbb{Z})$ are converted to the points on the circles.

We show the allowed region of $a$--$b$ in Fig.\,\ref{fig:ab}.
The magnitudes of $a$ and $b$ are found to be of order one for both NH and IH,
which is consistent with the conventional $A_4$ flavor model \cite{Shimizu:2011xg}.
It is noticed that the $(a,b)=(0,0)$ point is excluded.
That is to say, we need either singlet modular forms of $1'$ or $1''$
in order to reproduce the experimental data of leptons in Appendix B.
%%%%%%%%%%%%%%%%%%%
One naively expects to be $a\simeq b$
since $\Gamma_4\simeq S_4$ is broken to $A_4$ due to quantum effects
(anomaly) as discussed in section 3.
Obtained values of $a$ and $b$ deviate from $a\simeq b$ as seen in Eq.(\ref{sample-NH}) for NH while the desirable region of $a\simeq b$ exists as seen in Eq.(\ref{sample-IH}) for IH.
Thus, we should discuss the magnitude of the $S_4$ breaking beyond the naive expectation.
However, it is out of scope in this paper.
%%%%%%%%%%%%%%%%%%%%%%%%%%%%%%%%%%%%%%%%%%%%%%%

In our work, we take $a$ and $b$ to be real in a simple viable model.
Our predicted regions of $\delta_{CP}$ and $\langle m_{ee} \rangle$ are possibly enlarged if $a$ and $b$ are complex.
Whereas, it is worthwhile to discuss the case of real $a$ and $b$ because the case is attractive in the context of the generalized CP violation of modular-invariant flavor model \cite{Novichkov:2019sqv}.

%%%%%%%%%%%%%%%%%%%%%%%%%%%%%%%%%%%%%%%%%%%%%
%%%%%%%%%%%%%%%%%%%%%%%%%%%%%%%%%%%%%%%%%%%%
Finally, we also comment on numerical values on $\alpha$, $\beta$ and $\gamma$ of the charged lepton mass matrix.
These ratios are typically $\gamma/\beta={\cal O}(0.1)$ and  $\alpha/\beta={\cal O}(10^{-4})$ for the case of NH as seen in Eq.(\ref{sample-NH}).
The value of $\alpha$ is much smaller than $\beta$ and $\gamma$, on the other hand,
we need a mild hierarchy of ${\cal O}(0.1)$ between $\beta$ and  $\gamma$ although one may naively expect $\beta\sim \gamma$ as discussed in section 3.
Thus, the magnitude of the $S_4$ breaking is somewhat large beyond the naive expectation. 
For the case of IH, we need a strong hierarchy between $\beta$ and  $\gamma$ as seen in Eq.(\ref{sample-IH}).
Therefore, the IH case is not favored in our model.
%%%%%%%%%%%%%%%%%%%%%%%%%%%%%%%%%%%%%%%%%%%%%%%
%%%%%%%%%%%%%%%%%%%%%%%%%%%%%%%%%%%%%%%%%%%%

In our calculations, 
we take Yukawa couplings of charged leptons at the GUT scale $2\times 10^{16}$ GeV, where $v_u/v_d=2.5$ is taken as discussed in Appendix B.
However, we input the data of NuFIT 4.0 \cite{Esteban:2018azc} for three lepton mixing angles and neutrino mass parameters.
The renormalization group equation (RGE) effects of mixing angles and the mass ratio $\Delta m_{\rm sol}^2/\Delta m_{\rm atm}^2$ are negligibly small in the case of $\tan\beta=2.5$ even if IH of neutrino masses is considered (see Appendix B).
%%%%%%%%%%%%%%%%%%%%%%%%%%%%%%%%%%%%%%%%%%%%%%%%%%%%%

\section{Summary}

In the $S_4$ symmetry, the $Z_2$ subgroup can be anomalous, and then $S_4$ can be violated to $A_4$.
The $S_4$ symmetry is unique among  $S_3$, $A_4$, $S_4$, $A_5$ in the sense that  it can be violated by anomalies to another non-Abelian symmetry, $A_4$.
Starting with a $S_4$ symmetric Lagrangian at the tree level, the Lagrangian at the quantum level has only $A_4$ symmetry when $Z_2$ in $S_4$ is anomalous.
We have studied such a possibility that the $A_4$ flavor symmetry is originated from the  $S_4$ modular group.
Decomposing $S_4$ modular forms into  $A_4$ representations,
we have obtained the modular forms of two singlets, 
$\bf 1'$ and  $\bf 1''$, in addition to triplet, $\bf 3$ 
for $A_4$.
Using those modular forms,
we have succeeded in constructing the viable neutrino mass matrix through the Weinberg operator for both NH and IH of neutrino masses.
Our model presents a new possibility of flavor model with the modular symmetry. 

Indeed, we have obtained  an interesting prediction  of $\delta_{CP}$ for both NH and IH,
and their predictions also depend on the sum of neutrino masses. 
Hence, the observation of the sum of neutrino masses in the cosmology will provide a severe constraint to the flavor model.

Realistic mass matrices are realized in the parameter region with small $ {\rm Im}[\tau]$ as well as  large $ {\rm Im}[\tau]$.
If our four-dimensional field theory is originated from extra dimensional theory or superstring theory on a compact space, the volume of compact space is proportional to $ {\rm Im}[\tau]$.
Such volume of the compact space must be larger than the string scale.
For example, the volume of torus compactification is obtained by $(2\pi R)^2 {\rm Im}[\tau]$.
Thus, larger $2\pi R$ will be required for smaller $ {\rm Im}[\tau]$.

Furthermore, it is important how to derive the preferred values of $\tau$ in such compactified theory.
That is the so-called moduli stabilization problem.
However, that is beyond our scope.
We can study this problem elsewhere\footnote{
	Realistic results are obtained at some points of $\tau$ near edges of the $SL(2,{\mathbb Z})$ fundamental domain and domains transformed by $S$, $T$ and their products.
	The edges of the fundamental domain can be candidates for the minimum of the modulus potential.
	(See e.g.  Ref.\cite{Kobayashi:2016mzg} and its references therein.)}.

%%%%%%%%%%%%%%%%%%%%%%%%%%%%%%%%%%%%%%%%%%%%%%%%%%%%
%-------- acknowledgement -------%
\vspace{0.5cm}
\noindent
{\large\bf Acknowledgement}\\

This work is supported by  MEXT KAKENHI Grant Number JP19H04605 (TK), and 
JSPS Grants-in-Aid for Scientific Research  18J11233 (THT).
The work of YS is supported by JSPS KAKENHI Grant Number JP17K05418 and Fujyukai Foundation.

%-------- Appendix -------%
\appendix
\section*{Appendix}

\section{$S_4$ and $A_4$ representations}

The representations $S$ and $T$ of $\Gamma_4 \simeq S_4$ are given for the representations ${\bf 2}$ 
and ${\bf 3}'$ in section 2.
Here, we give other representations.
The generators $S$ and $T$ are represented by 
\begin{equation}
	\rho(S)=\frac13\left(
	\begin{array}{ccc}
		-1        & 2\omega^2 & 2 \omega  \\
		2\omega   & 2         & -\omega^2 \\
		2\omega^2 & -\omega   & 2         
	\end{array}\right), \qquad
	\rho(T)=
	\frac13\left(
	\begin{array}{ccc}
		-1        & 2\omega   & 2 \omega^2 \\
		2\omega   & 2\omega^2 & -1         \\
		2\omega^2 & -1        & 2\omega    
	\end{array}\right),
\end{equation}
on the $S_4$ ${\bf 3}$ representation, 
where $\omega=e^{i\frac{2}{3}\pi}$,
and 
\begin{equation}
	\rho(S) = \rho(T) =-1,
\end{equation}
for ${\bf 1}'$, while $\rho(S)=\rho(T)=1$ for ${\bf 1}$.

On the other hand, we take the generators of $A_4$ group
$\tilde S$ and $\tilde T$ for $\bf 3$ by using the $S$ and $T$ of the $S_4$ group as follows:
\begin{align}
	\begin{aligned}
	\rho(\tilde S)=\rho(T^2)=\frac{1}{3}
	\begin{pmatrix}
	-1 & 2        & 2      \\
	2  & -1       & 2      \\
	2  & 2        & -1     
	\end{pmatrix},
	\end{aligned}
	\qquad 
	\begin{aligned}
	\rho(\tilde T)=\rho(ST)=
	\begin{pmatrix}
	1  & 0        & 0      \\
	0  & \omega^2 & 0      \\
	0  & 0        & \omega 
	\end{pmatrix}. 
	\end{aligned}
\end{align}
Since the doublet ${\bf 2}$ of $S_4$ group is transformed by $\tilde{S}$ and $\tilde{T}$ as
\begin{align}
	\begin{aligned}
	\rho(\tilde S)=\rho(T^2)=
	\begin{pmatrix}
	1      & 0        \\
	0      & 1        \\
	\end{pmatrix},
	\end{aligned}
	\qquad 
	\begin{aligned}
	\rho(\tilde T)=\rho(ST)=
	\begin{pmatrix}
	\omega & 0        \\
	0      & \omega^2 \\
	\end{pmatrix},
	\end{aligned}
\end{align}
the doublet of $S_4$ can be decomposed into singlets of $A_4$ transformed as
\begin{align}
	\rho(\tilde{S})_{\bf 1'}=\rho(\tilde{S})_{\bf 1''}=1,\quad \rho(\tilde{T})_{\bf 1'}=\omega^2,\quad\rho(\tilde{T})_{\bf 1''}=\omega. 
\end{align}
In this base, the multiplication rule of the $A_4$ triplet is
\begin{align}
	\begin{pmatrix}
	a_1\\
	a_2\\
	a_3
	\end{pmatrix}_{\bf 3}
	\otimes 
	\begin{pmatrix}
	b_1\\
	b_2\\
	b_3
	\end{pmatrix}_{\bf 3}
	                                             & =\left (a_1b_1+a_2b_3+a_3b_2\right )_{\bf 1}                     
	\oplus \left (a_3b_3+a_1b_2+a_2b_1\right )_{{\bf 1}'}  \nonumber \\
	                                             & \oplus \left (a_2b_2+a_1b_3+a_3b_1\right )_{{\bf 1}''} \nonumber \\
	                                             & \oplus \frac13                                                   
	\begin{pmatrix}
	2a_1b_1-a_2b_3-a_3b_2 \\
	2a_3b_3-a_1b_2-a_2b_1 \\
	2a_2b_2-a_1b_3-a_3b_1 \\
	\end{pmatrix}_{{\bf 3}}
	\oplus \frac12
	\begin{pmatrix}
	a_2b_3-a_3b_2 \\
	a_1b_2-a_2b_1 \\
	a_3b_1-a_1b_3 \\
	\end{pmatrix}_{{\bf 3}\  } \ , \nonumber \\
	\nonumber \\
	{\bf 1} \otimes {\bf 1} = {\bf 1} \ , \qquad &                                                                  
	{\bf 1'} \otimes {\bf 1'} = {\bf 1''} \ , \qquad
	{\bf 1''} \otimes {\bf 1''} = {\bf 1'} \ , \qquad
	{\bf 1'} \otimes {\bf 1''} = {\bf 1} \  .
\end{align}

%\begin{align}
%	\begin{pmatrix}
%		a_1\\
%		a_2\\
%		a_3
%	\end{pmatrix}_{\bf 3}
%	\otimes 
%	\begin{pmatrix}
%		b_1\\
%		b_2\\
%		b_3
%	\end{pmatrix}_{\bf 3}
%	&=\left (a_1b_1+a_2b_3+a_3b_2\right )_{\bf 1} 
%	\oplus \left (a_2b_2+a_1b_3+a_3b_1\right )_{{\bf 1}'}  \nonumber \\
%	&\oplus \left (a_3b_3+a_1b_2+a_2b_1\right )_{{\bf 1}''} \nonumber \\
%	&\oplus \frac13
%	\begin{pmatrix}
%		2a_1b_1-a_2b_3-a_3b_2 \\
%		2a_2b_2-a_1b_3-a_3b_1 \\
%	2a_3b_3-a_1b_2-a_2b_1
%	\end{pmatrix}_{{\bf 3}}
%	\oplus \frac12
%	\begin{pmatrix}
%		a_2b_3-a_3b_2 \\
%			a_3b_1-a_1b_3 \\
%	a_1b_2-a_2b_1
%	\end{pmatrix}_{{\bf 3}\  } \ , \nonumber \\
%	\nonumber \\
%	%%
%	{\bf 1} \otimes {\bf 1} = {\bf 1} \ , \qquad &
%	{\bf 1'} \otimes {\bf 1'} = {\bf 1''} \ , \qquad
%	{\bf 1''} \otimes {\bf 1''} = {\bf 1'} \ , \qquad
%	{\bf 1'} \otimes {\bf 1''} = {\bf 1} \  .
%\end{align}

More details are shown in the review~\cite{Ishimori:2010au,Ishimori:2012zz}.

\section{Input data}
%%%%%%%%%%%%%%%%%%%%%%%%%%%%%%%%%%%%%%%%%%%%%%%%%%%%%%%%%%%%%%%%%%%%%%%
%%%%%%%%%%%%%%%%%%%%%%%%%%%
We input charged lepton masses in order to constrain the model parameters.
We take Yukawa couplings of charged leptons 
at the GUT scale $2\times 10^{16}$ GeV,  where $\tan\beta=2.5$ is taken
\cite{Criado:2018thu,Antusch:2005gp,Antusch:2013jca,Bjorkeroth:2015ora}:
\begin{eqnarray}
	y_e=(1.97\pm 0.02) \times 10^{-6}, \quad 
	y_\mu=(4.16\pm 0.05) \times 10^{-4}, \quad 
	y_\tau=(7.07\pm 0.07) \times 10^{-3},
\end{eqnarray}
where lepton masses are  given by $m_\ell=\sqrt{2} y_\ell v_H$ with $v_H=174$ GeV.
%%%%%%%%%%%%%%%%%%%%%%%%%%%%%%%%%%%%%%%%%%%%%%%%%%%%%%%%%%%%%%%%%%%%%%%%%
We also use the following lepton mixing angles and neutrino mass parameters in Table 2 given by NuFIT 4.0 \cite{Esteban:2018azc}.
The RGE effects of mixing angles and the mass ratio $\Delta m_{\rm sol}^2/\Delta m_{\rm atm}^2$
are negligibly small in the case of $\tan\beta=2.5$ for both NH and IH as seen in Appendix E of Ref.\,\cite{Criado:2018thu}.
%%%%%%%%%%%%%%%%%%%%%%%%%%%%%%%%%%%%%%%%%%%%%%%%%%%%%%%%%%%%%%%%%%%%%%%%%%%%%%
\begin{table}[h]
	\begin{center}
		\begin{tabular}{|c|c|c|}
			\hline 
			\rule[14pt]{0pt}{0pt}
			\  observable \&$3\,\sigma$ range for NH & $3\,\sigma$ range for IH &                      \\
			\hline 
			\rule[14pt]{0pt}{0pt}
			%&& \\
			$\Delta m_{\rm atm}^2$& \ \   \ \ $(2.431$--$ 2.622) \times 10^{-3}{\rm eV}^2$ \ \ \ \
			&\ \ $- (2.413$--$2.606) \times 10^{-3}{\rm eV}^2$ \ \  \\
			%&& \\
			\hline 
			\rule[14pt]{0pt}{0pt}
			%&& \\
			$\Delta m_{\rm sol }^2$& $(6.79$--$ 8.01) \times 10^{-5}{\rm eV}^2$
			& $(6.79$--$ 8.01)  \times 10^{-5}{\rm eV}^2$ \\
			%&& \\
			\hline 
			\rule[14pt]{0pt}{0pt}
			%&& \\
			$\sin^2\theta_{23}$                      & $0.428$--$ 0.624$        & $0.433$--$ 0.623$    \\
			%&& \\
			\hline 
			\rule[14pt]{0pt}{0pt}
			%&& \\
			$\sin^2\theta_{12}$                      & $0.275$--$ 0.350$        & $0.275$--$ 0.350$    \\
			%&& \\
			\hline 
			\rule[14pt]{0pt}{0pt}
			%&& \\
			$\sin^2\theta_{13}$                      & $0.02044$--$ 0.02437$    & $0.02067$--$0.02461$ \\
			%&& \\
			\hline 
		\end{tabular}
		\caption{The $3\,\sigma$ ranges of neutrino  parameters from NuFIT 4.0
			for NH and IH 
			\cite{Esteban:2018azc}. 
		}
		\label{DataNufit}
	\end{center}
\end{table}

%%%%%%%%%%%%%%%%%%%%%%%%%%%%%%%%%%%%%%%%%

%\newpage

\end{document}